\documentclass[12pt,a4paper]{article}
\usepackage[latin1]{inputenc}
\usepackage[T1]{fontenc}
\usepackage{amsmath}
\usepackage{amsfonts}
\usepackage[english]{babel}
\begin{document}
\title{Casimir bag energy in the stochastic approximation to the pure QCD vacuum}
\author{C. D. Fosco$^a$ and L. E. Oxman$^b$\\
  {\normalsize\it $^a$Centro At\'omico Bariloche and Instituto Balseiro}\\
  {\normalsize\it Comisi\'on Nacional de Energ\'{\i}a At\'omica}\\
  {\normalsize\it R8402AGP Bariloche, Argentina.}\\
  {\normalsize\it $^b$ Instituto de F\'{\i}sica, Universidade
Federal Fluminense}\\
  {\normalsize\it Av. Litor\^anea S/N, Boa Viagem,}\\
  {\normalsize\it Niter\'oi, RJ 24210-340, Brazil.}}
\date{}
\maketitle
\begin{center}
\vspace{0.6cm}
\end{center}
\begin{abstract}
We study the Casimir contribution to the bag energy coming from gluon
field fluctuations, within the context of the stochastic vacuum model
(SVM) of pure QCD. After formulating the problem in terms of the generating 
functional of field strength cumulants, we argue that the
resulting predictions about the Casimir energy are compatible with the
phenomenologically required bag energy term.
\end{abstract}
\maketitle
\section{Introduction}

In order to adjust hadron spectroscopy, the MIT bag model requires a
bag energy term of the form $-Z/R$ with a positive $Z$ of order
one~\cite{dejjk}.
One of the main contributions to this energy is naturally ascribed to the Casimir 
energy of the fields involved.
However, due to the daunting difficulty posed by Casimir energy
calculations in nonabelian gauge field theories, this problem has only
been approached under rather strong simplifying assumptions.  Indeed,
vacuum effects coming from the pure Yang-Mills sector have been
computed by considering a simplified description, which treats gluon
modes as a collection of free, photon-like gauge fields.  This assumption
(usually justified by invoking asymptotic freedom) leads, however, to
an energy with the wrong sign: $\sim +0.7/R$. Thus the effect becomes
{\em repulsive}, instead of the attractive one needed to comply with
phenomenology~\cite{boyer}-\cite{milton3}. For a discussion of the bag
model, as well as of the current status of Casimir calculations in the bag, 
see, for example, refs.~\cite{Milton} and~\cite{Don}.

Neither the inclusion of the (rather small) fermionic Casimir energy
contribution $\sim - 0.02/R$ due to three light quark modes~\cite{milton2},
nor the (relevant) effect of the quarks' center of mass quantum
fluctuations, which is attractive, $\sim -Z_{cm}/R$, \mbox{$Z_{cm}\sim
0.6-0.8$}, are sufficient to reach the required value of $Z$ (see the
discussion in~\cite{Milton} and \cite{DJWL}).

It has been pointed out that a resolution to this problem could be
associated with the intrinsically nonabelian nature of the Yang-Mills
theories, and their ensuing nonperturbative effects. This is certainly
plausible, although it is not at all evident how to include those
effects in a tractable way.

In this article, we present an approach that enables one to take partly
into account the nonperturbative dynamics of QCD into the
calculation of the Casimir energy, and argue that its predictions are
phenomenologically sound.

We first show how the Casimir energy in pure QCD can be written in
terms of a subclass of gauge invariant correlators, namely, the
shifted field-strength correlators. Lattice simulations show~\cite{latt} that 
these correlators exhibit Gaussian dominance, so that 
they can be described in terms of the so called stochastic vacuum
model (SVM) of pure QCD~\cite{svm} (for a review see refs.~\cite{rep} 
and \cite{Otto}). This model can be parametrized in
terms of just two structure functions, obtained by fitting the lattice
data, and it consistently describes short distance perturbative
aspects, such as asymptotic freedom, as well as long distance
nonperturbative ones, such as confinement.

In order to relate Casimir energy calculations to correlators of the
stochastic model, we will use a framework where the boundary conditions are
imposed by means of auxiliary fields living on the bag boundary.  In this
way, we are able to show that the lattice-adjusted SVM parameters imply
that the simple model that computes Casimir energies treating gluon modes
as massless photon-like gauge fields is not reliable; rather, a scenario
with a suppressed Casimir energy due to gapped nonperturbative effective
modes is favoured. Thus, a physical picture arises where the
suppression of the gluon Casimir energy leaves room for the effect of
center of mass quantum fluctuations to dominate the $R^{-1}$
contribution to the bag energy.

The structure of this paper is as follows: in
section~\ref{sec:abelian}, as a preparation for the nonabelian case,
we use the approach of~\cite{kardar} to construct the path integral
for the Casimir energy for a confined Abelian gauge field with
magnetic boundary conditions, in terms of the field-strength
correlation function. In section~\ref{sec:nonabelian}, the same
approach is followed for a nonabelian gauge field within the context
of the SVM. In section~\ref{sec:combined}, we show that the same
vacuum energy that one obtains from the SVM may be alternatively
understood as due to a pair of fields, one of them vectorial and the
other pseudovectorial, both endowed with non-standard kinetic terms.  This
parametrization is then applied to infer the main properties of the
SVM Casimir energy.

\noindent In section~\ref{sec:conclusions}, we present our conclusions.
\section{Confined Abelian field}\label{sec:abelian}
Let us first consider an Abelian gauge field $A_\mu$ in $3+1$
dimensions, confined to a spatial region ${\mathcal V}$ whose boundary
is a regular (static) surface $\Sigma$. On that surface, we shall
consider boundary conditions corresponding to a perfect conductor of
magnetic charges.

We shall assume that $\Sigma$ can be described in parametric form,
\begin{eqnarray}
{\mathbf r}: U &\longrightarrow& {\mathbb R}^3 \nonumber\\
 (\zeta^1, \zeta^2) &\longrightarrow& 
{\mathbf r}(\zeta^1,\zeta^2)\;,
\end{eqnarray}
where $U$ is a finite region of ${\mathbb R}^2$, the domain of the
parameters $\zeta^1,\, \zeta^2$. In order the impose the boundary
conditions, it is convenient to introduce first some notation and
conventions. Since the boundary conditions are static, they shall hold
on the world-volume $\Gamma = {\mathbb R} \times \Sigma$ swept by
$\Sigma$ in the course of its (trivial) time evolution.  This
world-volume can be parametrized by a function $r_\mu$ ($\mu =
0,1,2,3$), depending on \mbox{$\zeta^1,\, \zeta^2$} as well as on an
extra parameter $\zeta^0$:
\begin{eqnarray}
r_\mu : {\mathbb R} \times U &\longrightarrow& {\mathbb R}^4
\nonumber\\
 (\zeta^0, \zeta^1, \zeta^2) &\longrightarrow& 
 r_\mu (\zeta)\;,
\end{eqnarray}
where \mbox{$r_0(\zeta) = \zeta_0$} and \mbox{$r_i(\zeta) = r_i
(\zeta_1,\zeta_2)$}.

Regarding the `target' space, we shall work in Euclidean spacetime,
with coordinates $x_\mu$, \mbox{$\mu = 0,1,2,3$}, and a metric which
is the identity matrix (no special meaning will be given to the
position -up or down- of space and spacetime indices).  Letters from
the beginning of the Greek alphabet run from $0$ to $2$, those from
the middle do so from $0$ to $3$, while Roman ones can take values
from $1$ to $3$. Besides, $a,b,c,\ldots$ will be reserved for surface
coordinates, and therefore may assume the values $1$ or $2$.

At any point of $\Gamma$, we have the three tangent vectors
$t^\mu_\alpha = \frac{\partial r^\mu}{\partial \zeta^\alpha}$, which
are not necessarily orthogonal but certainly linearly independent
vectors~\footnote{At least under the assumption that $\Sigma$ is a
regular surface. They may be normalized, but that step shall not be
necessary.}. Of course, $t_1$ and $t_2$ are tangent to $\Sigma$, at
all times.

Besides, we note that each component $t^\mu$ of $t$ transforms as a
covariant vector under reparametrizations:  $t'_\alpha (\zeta') =
\frac{\partial\zeta^\beta}{\partial\zeta'^\alpha} \, t_\beta (\zeta)$.

Finally, we shall also make use of ${\mathbf n}$, the (outer) normal to
$\Sigma$.  At any point ${\mathbf r}(\zeta)$ it can be easily obtained
as the vector product of the two tangent vectors $t_1$ and $t_2$:
${\mathbf n} = {\mathbf t}_1 \times {\mathbf t}_2$.  This vector may
be easily normalized by taking into account the fact that:
\begin{equation}
{\mathbf n}^2(\zeta) \;=\; \big({\mathbf t}_1 (\zeta)\,\times\, {\mathbf
t}_2(\zeta) \big)^2
\;=\, g_\Sigma(\zeta) \;,
\end{equation}
where $g_\Sigma \equiv \det(g_{ab})$, and $g_{ab}$ is the
induced metric on $\Sigma$:
\begin{equation}
g_{ab}(\zeta) \;=\; t^\mu_a (\zeta) t^\mu_b (\zeta) \;;
\end{equation}
thus, the unit normal $\mathbf{\hat{n}}$ is simply: $\mathbf{\hat{n}}
\equiv \frac{\mathbf n}{|{\mathbf n}|} = \frac{1}{\sqrt{g_\Sigma}} \,
{\mathbf n}$.  As a matter of fact, since the time evolution of the
boundary is trivial, $g_\Gamma$, the determinant of the induced metric
on $\Gamma$, coincides with $g_\Sigma$.
 
Regarding the transformation properties of $\mathbf{\hat{n}}$, one
easily sees that it behaves as a pseudo-scalar under surface
reparametrizations.

Now we introduce the magnetic boundary conditions: they amount to
surrounding the region ${\mathcal V}$ with a perfect magnetic
conductor, so that the tangential component of ${\mathbf B}$ at every
point of $\Sigma$, and for every time vanishes:
\begin{equation}\label{eq:bcond1}
{\mathbf {\hat n}}(\zeta) \times {\mathbf B}\big(r_\mu(\zeta)\big) 
\;=\; {\mathbf 0}\;,
\end{equation}
or, in terms of $F_{\mu\nu}$,
\begin{equation}\label{eq:bcond2}
F_{i j}\big( r_\mu(\zeta) \big) {\hat n}^j(\zeta) \;=\; 0 \;,\;\;\;
i=1,2,3\,.
\end{equation}
Note that they do not amount to three independent conditions, since
they satisfy the condition: \mbox{$\hat{n}_i F_{i j}\hat{n}_j=0$}. Thus,
(\ref{eq:bcond2}) may be written equivalently as:
\begin{equation}\label{eq:bcond3}
\eta_a \;\equiv \;t_a^i \, F_{ij} \, \hat{n}^j 
\;=\;0\;\;,\;\;\;\; a = 1, 2\;,
\end{equation}
where only the two independent constraints, $\eta_1$ and $\eta_2$
appear (note that $t^i_0 = 0$, thus there is no other constraint).  

Let us now consider the Euclidean path integral that result from
imposing the previously introduced constraints into the vacuum
transition amplitude for the Maxwell field. To that end, we shall
define ${\mathcal Z}_{mag}$, the partition function corresponding to a
magnetic charge conductor, since it is more similar to the case we
will need to deal with in the nonabelian case:
\begin{eqnarray}\label{eq:defzmag}
{\mathcal Z}_{mag} &\equiv& \int [{\mathcal D}A]_{mag} \;
e^{- S_g [A]}\nonumber\\
&=& {\mathcal N} \, \int {\mathcal D}A \, \prod_{a=1}^2 \delta[
\eta_a] \; 
e^{- S_g [A]}\;,
\end{eqnarray}
where $S_g$ is the gauge field action (including gauge fixing), and
$[{\mathcal D}A]_{mag}$ is the integration measure for the gauge
field, assuming it satisfies the corresponding boundary conditions on
$\Sigma$.  Following~\cite{kardar}, those conditions have been
imposed, on the second line, by means of two $\delta$ functionals. The
integration measure for $A_\mu$ becomes then unconstrained.

It is convenient to exponentiate the $\delta$ functionals, by means of
two Lagrange multipliers, denoted by $\phi^a(\zeta)$.  It has been
shown that the exponentiated $\delta$-functionals have to be invariant
under reparametrizations of the surface~\cite{kardar}.  Thus, it is
immediate to see that the proper expression is:
\begin{equation}
\prod_{a=1}^2 \delta[\eta_a] \;=\; \int {\mathcal D}\phi 
\, e^{ - S_\delta [\phi, A]} 
\end{equation}
where
\begin{equation}
S_\delta [\phi, A] \;=\; -\, i\, \int d^3 \zeta \, 
\sqrt{g_\Gamma(\zeta)} \, 
\phi^a (\zeta)  \eta_a(\zeta) \;. 
\end{equation}

Reparametrization invariance of $S_\delta$ is easily verified; indeed,
$d^3\zeta \, \sqrt{g_\Gamma(\zeta)}$ is invariant by itself, while
$\eta_a$ is a covariant vector, whose change is compensated by the
variation of the contravariant vector $\phi^a$.  Note that the
\mbox{$\sqrt{g_\Gamma(\zeta)} = \sqrt{g_\Sigma(\zeta)}$} factor may be 
cancelled with a like one in the unit normal; thus:
\begin{eqnarray}
S_\delta [\phi, A] &=& -\, i\, \int d^3 \zeta \, 
\phi^a(\zeta)  t_a^i(\zeta) F_{ij}\big(r_\mu(\zeta)\big) n^j
(\zeta)\nonumber\\
&=& -\, i\, \int d^3 \zeta \,\phi^a(\zeta)  \partial_a r^i(\zeta) 
F_{ij}\big(r_\mu(\zeta)\big) \epsilon^{jkl} \partial_1 r^k(\zeta)
\partial_2 r^l(\zeta) \;. 
\end{eqnarray} 
This term may be regarded as an interaction between the gauge field
and a `current' $J^\Sigma$ due to the boundary conditions: 
\begin{equation}
S_\delta \;=\; - \, \frac{i}{2} \, \int d^4x \, J^\Sigma_{kl}(x) F_{kl}(x)
\end{equation}
with
\begin{equation}\label{eq:defjkl}
J^\Sigma_{kl}(x) \;=\; \int d^3\zeta \, \phi^a (\zeta)
\, \delta^{(4)}\big( x_\mu - r_\mu (\zeta) \big) \, 
\big[ (\partial_a r_k)  n_l - (\partial_a r_l) n_k \big](\zeta) \;.
\end{equation}

Thus, in terms of ${\mathcal Z}(J_{\mu\nu})$, the generating functional of
field strength correlators:
\begin{equation}
{\mathcal Z}(J_{\mu\nu}) \;\equiv\; e^{-{\mathcal W}(J_{\mu\nu})}
\,=\,\frac{\int {\mathcal D}A \; e^{- S_g (A) + \frac{i}{2} \int d^4x
J_{\mu\nu} F_{\mu\nu}}}{\int {\mathcal D}A \;e^{- S_g (A)}}\;,
\end{equation}
${\mathcal Z}_{mag}$ may be written as:
\begin{equation}
{\mathcal Z}_{mag} \;=\;\int {\mathcal D}\phi \, \exp[- {\mathcal
W}(J_{\mu\nu})] \Big|_{J = J^\Sigma} \;,
\end{equation}
where $J_{\mu\nu}^\Sigma$ agrees with (\ref{eq:defjkl}) when both
indices are spatial, and is assumed to vanish when one of them equals
zero. Of course, the integral is Gaussian, and one is able to write
the result:
\begin{equation}
{\mathcal W}(J_{\mu\nu}) \;=\; \frac{1}{8} \, \int d^4x \int d^4y \,
J_{\mu\nu}(x) \langle F_{\mu\nu}(x) F_{\rho\sigma}(y) \rangle
J_{\rho\sigma}(y)\;,
\end{equation}
where:
\begin{eqnarray}
\langle F_{\mu\nu}(x) F_{\rho\sigma}(y) \rangle 
\;\equiv\;  \Delta_{\mu\nu,\rho\sigma} (x-y)
\end{eqnarray}
with
\begin{equation}
\Delta_{\mu\nu,\rho\sigma} (x) \;=\; \int \frac{d^4k}{(2\pi)^4} e^{i k
\cdot x}\, {\tilde \Delta}_{\mu\nu,\rho\sigma} (k) 
\end{equation}
and
\begin{equation}\label{eq:ffk}
{\tilde \Delta}_{\mu\nu,\rho\sigma} (k) \;=\; - \frac{1}{k^2}
\Big( \delta_{\nu\sigma} \, k_\mu k_\rho 
+ \delta_{\mu\rho} \, k_\nu k_\sigma -  
\delta_{\mu\sigma} \, k_\nu k_\rho - \delta_{\nu\rho} \, 
k_\mu k_\sigma \Big)\;.
\end{equation}

The coordinate-space version of (\ref{eq:ffk}) may be expressed as
follows:
\begin{equation}\label{eq:ffk1}
\Delta_{\mu\nu,\rho\sigma} (z) \;=\; 
\frac{1}{2} \Big[\partial_\nu
(z_\sigma \delta_{\mu\rho} 
- z_\rho \delta_{\mu\sigma} )
+ \partial_\mu (z_\rho \delta_{\nu\sigma} 
- z_\sigma \delta_{\nu\rho} )\Big]
D_0(z^2)\;,
\end{equation}
where
\begin{equation}
z \equiv x-y\;,\;\;\; 
D_0(z^2) \;\equiv\; \frac{1}{\pi^2} \frac{d}{dz^2} \frac{1}{z^2}
\;=\; - \frac{1}{\pi^2 z^4}\;.
\end{equation}
The vacuum energy $E_0$ in the presence of the boundary $\Sigma$ is then
given by:
\begin{equation}
E_0 \;=\; - \, \lim_{T\to \infty} \, \frac{1}{T} \ln {\mathcal Z}_{mag}\;,
\end{equation}
where $T$ denotes the extension of the Euclidean time coordinate:
$|x_0| \leq T/2$.

\section{Confined nonabelian gauge field}\label{sec:nonabelian}
For a region surrounded by a perfect conductor of chromomagnetic charges, 
we have at the boundary the condition:
\begin{equation}\label{eq:chrcond2}
{\bf B}\times {\bf \hat{n}}={\bf 0}
\end{equation} 
where ${\bf B} \equiv {\bf B}^u T_u$ and $T_u$ are the
generators of the Lie group. The index $u$ runs from $1$ to $N$, the
dimension of the adjoint representation (i.e., $N=8$, for $SU(3)$).
In terms of the nonabelian field strength,
\begin{equation}\label{eq:chrcond1}
 F_{i j}(r_\mu(\zeta)) \hat{n}_j(\zeta) \,=\, 0 \;.
\end{equation}
As the constraints (\ref{eq:chrcond1}) are homogeneous, they can also be
written in the equivalent form
\begin{equation}\label{scm}
F_{i j}(y, x ;{\mathcal C}) 
\hat{n}_j(\zeta) \,=\, 0 \;,\;\; \forall x \in \Sigma\;,
\end{equation}
obtained by covariantly shifting the field strength $F_{i j}$ from an
arbitrary (fixed) reference point $y$.  $F_{\mu \nu}(y,x;{\mathcal C})$
is the shifted field strength, defined by, 
\begin{equation}
F_{\mu \nu}(y,x;{\mathcal C}) \,=\, V(y,x;{\mathcal C})
\, F_{\mu\nu}(x)\, V^{-1}(y,x;{\mathcal C}),
\end{equation}
where $V$ is the parallel transporter, from $x$ to $y$, along a path
${\mathcal C}$,
\begin{equation}
V(y,x;{\mathcal C})\,=\,{\mathcal P} 
\exp\left( -ig\int_{\mathcal C} dz^\mu A_\mu(z) \right)\;, 
\end{equation}
and ${\mathcal P}$ is the path ordering operator.

The Casimir energy obtained by imposing either the constraints
(\ref{eq:chrcond1}) or the equivalent ones (\ref{scm}) should
be the same, since they differ by a gauge transformation~\footnote{
This could be verified, for example, by means of an appropriate
transformation of the corresponding Lagrange multipliers.}.

Thus we will impose the constraints,
\begin{equation}
\eta^u_a\;\equiv \; t_a^i {\hat n}^j F^u_{i j}(x,y;{\mathcal
C})\;=\; 0 \;.
\end{equation}
Introducing, as in the previous section, Lagrange multipliers
$\phi^a_u$, we obtain the representation,
\begin{equation}
Z_{mag}\;=\;\int [{\cal D}\phi]\, e^{-S(\phi)}
\end{equation}
defined as in the Abelian case, but with $\phi$, being a pair of fields
in the adjoint representation, and with a generating functional that
now corresponds to the nonabelian case:
\begin{equation}
{\mathcal W} (J)= -\ln {\mathcal Z} (J) \;,
\end{equation}
with
\begin{equation}\label{naZ}
{\mathcal Z}(J)\,\equiv\,\int [{\cal D}A]\, e^{ -S_g(A) \,+\, 
\frac{i}{2}\int d^4x\, F^u_{\mu \nu}(y,x;\, {\mathcal C}) J_u^{\mu
\nu}(x)} \;,
\end{equation}
where $S_g(A)$ is the gauge-fixed Yang-Mills action.

The vacuum energy $E_0$, defined as in the Abelian case, must be gauge
invariant. This fact, naturally expected from the gauge invariance of
the boundary conditions at $\Sigma$, may be verified explicitly as
follows: under a gauge transformation associated with the group
transformation $U(x)$, the shifted field strength transforms in the
adjoint representation, with the induced group rotation evaluated at
$y$.  This change can be compensated by a parallel transport (also in
the adjoint representation) of the Lagrange multipliers to the point
$y$.  Being that a unitary transformation, no non-trivial Jacobian is
generated.

We also note that, had an Abelian theory been considered, we would
have $F_{\mu \nu}(y,x;{\mathcal C})=F_{\mu \nu}(x)$ and this
calculation would have reduced to the one performed in the previous
section.

Returning to the general nonabelian case, taking functional derivatives of
${\mathcal Z}(J)$ at $J=0$, we may generate field strength correlators,
\begin{equation}
\langle F^{u_1}_{\mu_1 \nu_1}(y,x_1;{\mathcal C}_1)\dots F^{u_n}_{\mu_n
\nu_n}(y,x_n;{\mathcal C}_n) \rangle 
\end{equation}
which, due to colour conservation satisfy:
\begin{equation}
\langle F^{u}_{\mu_1 \nu_1}(y,x_1;{\mathcal C}_1) \rangle \,=\, 0\;,
\end{equation}
and
\begin{equation}
\frac{g^2}{4\pi}\langle F^{u_1}_{\mu \nu}(y,x;{\mathcal C}) 
F^{u_2}_{\rho \sigma}(y,x';{\mathcal C}') \rangle=
\frac{1}{4}\delta^{u_1 u_2}F_{\mu \nu \rho \sigma}(x,x',y;{\mathcal
C},{\mathcal C}')\;.
\label{bl}
\end{equation}
It can also be shown that the correlator in (\ref{bl}) may only depend on
${\mathcal C} - {\mathcal C}'$, the curve connecting $x$ to $x'$. 

To proceed, we recall that the SVM is based on the strong assumption
that this correlator does not depend at all on ${\mathcal
C}$ and ${\mathcal C}'$. Moreover, the expectation value of the product of any number of
shifted $F$'s vanishes, while the product of an even number can be
decomposed in terms of the sum of the different products that can be
formed with just two field correlators.  In other words, the
assumptions of the SVM are tantamount to saying that the quantum
fluctuations are controlled by an essentially Gaussian measure.  To
define the model, it is then sufficient to specify a {\em quadratic\/}
${\mathcal W}(J)$ functional, with the structure:
\begin{equation}
{\mathcal Z}(J)\;=\; e^{- \frac{\pi}{8 g^2}\int d^4x\, d^4x'\,
J_u^{\mu \nu}(x) F_{\mu \nu ;\rho\sigma}(x,x')J_u^{\rho \sigma}(x')}\,.
\end{equation}
As a consequence,
\begin{equation}
{\mathcal Z}_{mag}\,=\,\left[ \int {\cal D}\phi \, 
e^{- \frac{\pi}{8 g^2}\int d^4x\, d^4x'\, J^\Sigma_{\mu \nu}(x) 
F_{\mu \nu ;\rho \sigma}(x,x')J^\Sigma_{\rho \sigma}(x')}\right]^N \;,
\label{Zmag}
\end{equation}
where now $\phi$ refers to just a single pair of fields $\phi^a$
(without any colour index).

The  correlator for two shifted field strengths is frequently
parametrized as follows~\footnote{For a recent review see, for example,
ref.~\cite{rep}.}:
\begin{eqnarray}
F_{\mu \nu ; \rho \sigma}(x,x')&=& 
(\delta_{\mu\rho}\delta_{\nu \sigma}-\delta_{\mu \sigma}\delta_{\nu
\rho})\kappa D(-z^2) \nonumber\\
&+&\frac{1}{2}\left[\partial_\nu(z_\sigma \delta_{\mu
\rho}-z_\rho \delta_{\mu \sigma})
+\partial_\mu(z_\rho \delta_{\nu
\sigma}-z_\sigma \delta_{\nu \rho})\right](1-\kappa) D_1(-z^2)\nonumber \\
\label{param}
\end{eqnarray}
where $z\equiv x-x'$. 

Then, within the SVM, the Casimir energy is just $N$ times the one
computed in some kind of generalized Abelian theory, namely, the one
which comes from the modified kernel $F_{\mu \nu ;\rho \sigma}$. Note,
in particular, that it contains a $D$-term, an object that is not
present in the Abelian case (because of the absence of magnetic
monopoles). This new term comes from the nonabelian character of the theory, 
and can be associated with the dual superconductor scenario~\cite{thooft} 
of the QCD vacuum, representing the effects of a chromomagnetic monopole condensate;
it is crucial to derive the confining linear potential between heavy charges. 
The second term (Abelian part) contains the modified nonlocal kernel $D_1$, where perturbative 
effects dominate. We will discuss later specific fittings of $D$ and $D_1$ which are obtained from 
lattice data.

\section{Field representation of the Casimir energy}\label{sec:combined}
An equivalent functional integral representation for the Casimir
energy, previously expressed in terms of shifted field strength
correlators, can be constructed in terms of Gaussian fields, albeit
with non standard propagators.
We have found that the most economical
parametrization is in terms of a vector field $A_\mu$ and a
pseudo-vector field $\Phi_\mu$:
\begin{equation}
F_{\mu \nu} = \partial_\mu A_\nu-\partial_\nu A_\mu+\epsilon_{\mu \nu
\rho \sigma}\partial_\rho \Phi_\sigma \;,
\label{Fmunu}
\end{equation}
where the new $\Phi$-dependent term represents the monopole sector of
nonperturbative Yang-Mills. Similar field representations have been considered to discuss 
Wilson loops, also in the context of the method of field strength 
correlators~\cite{antonov}. 

Then the $\langle F F\rangle$ correlation function is now given by:
\begin{eqnarray}
\lefteqn{\langle F_{\mu \nu}(x) F_{\rho \sigma}(y)\rangle =}\nonumber
\\ &=& \langle(\partial_{\mu} A_{\nu}-\partial_{\nu} A_{\mu}
)(\partial_{\rho} A_{\sigma}-\partial_{\sigma} A_{\rho})\rangle
+\epsilon_{\mu \nu \mu' \nu'}\epsilon_{\rho \sigma \rho' \sigma'}
\partial_{\mu'}\partial_{\rho'} \langle
\Phi_{\nu'}(x)\Phi_{\sigma'}(y)\rangle \nonumber\\ &\equiv& \hat{O}_{\mu
\nu,\rho \sigma} (G-S)' + I_{\mu \nu,\rho\sigma}\,
\partial^2\, S\; ,\nonumber \\
\end{eqnarray}
where $\langle A_\mu (x) \Phi_\nu (y) \rangle = 0$,
\begin{equation}
\langle A_\mu (x) A_\nu (y) \rangle =\delta_{\mu \nu} G(z^2)
\makebox[.3in]{,}
\langle \Phi_\mu (x) \Phi_\nu (y) \rangle =\delta_{\mu \nu} S(z^2) \;.
\end{equation}
The prime represents derivative with respect to $z^2$, and we used the
notation:
\begin{equation}
I_{\mu \nu\, ,\, \rho \sigma}\equiv (\delta_{\mu\rho}\delta_{\nu
\sigma}-\delta_{\mu\sigma}\delta_{\nu\rho})\;,
\end{equation}
\begin{equation}
\hat{O}_{\mu \nu\, ,\, \rho \sigma}= 2\{
\partial_{\nu}[\delta_{\mu \rho} z_{\sigma}-\delta_{\mu \sigma}
z_{\rho}] +\partial_{\mu}[\delta_{\nu \sigma} z_{\rho}-\delta_{\nu
\rho} z_{\sigma}]\}\;.
\end{equation}
Then, to reproduce the lattice parametrization of (\ref{param}), one
can make the identifications,
\begin{equation}
\partial^2\, S= \kappa D
\makebox[.5in]{,}
G'-S'=(1-\kappa) D_1/4\; ,
\label{H}
\end{equation}
which can be solved by imposing the following equations,
\begin{equation}
\tilde{S}=-\kappa \frac{\tilde{D}}{k^2}
\makebox[.3in]{,}
\tilde{G}=-\kappa \frac{\tilde{D}}{k^2}+(1-\kappa)\tilde{H}
\makebox[.3in]{,}
H'=D_1/4\;.
\label{ident}
\end{equation}
In other words, the (only) nontrivial propagators are,
\begin{eqnarray}
\langle A_\mu (x) A_\nu (y) \rangle &=& -\int \frac{d^4k}{(2\pi)^4}\,
\left[\kappa \tilde{D}+(1-\kappa)k^2\tilde{H}\right] \frac{\delta_{\mu
\nu}}{k^2} e^{ik\cdot(x-y)} \nonumber\\ 
\langle \Phi_\mu (x) \Phi_\nu (y) \rangle &=& -\int
\frac{d^4k}{(2\pi)^4}\; \left[ \kappa \tilde{D}\right] \;
\frac{\delta_{\mu\nu}}{k^2} e^{ik\cdot(x-y)}\; .
\end{eqnarray}
We see that $\tilde{D}$ and $k^2\tilde{H}$ became associated with
`structure functions' for the propagators of the $A_\mu$ and
$\Phi_\mu$ fields.

Then, the partition function (\ref{Zmag}) can be also represented by, ${\mathcal
Z}_{mag}={\mathcal Z}_{A,\Phi}^N$, where
\begin{equation}
{\mathcal Z}_{A,\Phi}\,=\, \int {\cal D}\phi {\cal D}A {\cal D}\Phi \, 
e^{ -S_e[A]-S_m[\Phi]+\frac{i}{2}\int d^4 x\, J^\Sigma_{\mu \nu}(\partial_\mu
A_\nu-\partial_\nu A_\mu+\epsilon_{\mu \nu \rho \sigma}\partial_\rho
\Phi_\sigma) }\; .
\end{equation}
Path integrating over the Lagrange multipliers $\phi$, we get the
constraints,
\begin{equation}
F_{ij}n^j=(\partial_i A_j-\partial_j A_i+\epsilon_{i j \rho \sigma}\partial_\rho \Phi_\sigma)n^j=0\; ,
\label{bc}
\end{equation}
namely, the Yang-Mills fields Casimir energy, due to Gaussian dominance
in the shifted field strength correlators, can be written as $N$ times
the one obtained for a field theory with action $S_e[A]+S_m[\Phi]$.
Here, both $S_e[A]$ and $S_m[\Phi]$ are (typically) nonlocal versions
of Maxwell-like actions; for instance, $S_m[\Phi]\sim \int d^4x\, G_{\mu
\nu}\big[{\tilde D}^{-1}(\partial^2)\big]G_{\mu \nu}$, with
$G_{\mu\nu}=\partial_\mu \Phi_\nu -\partial_\nu \Phi_\mu$.  We shall
assume that a Lorentz gauge-fixing condition for $A_\mu$ and
$\Phi_\mu$ is implicitly included in the functional integral measure.

Then, we have,
\begin{equation}
{\mathcal Z}_{A,\Phi}\,=\, \int [{\cal D}A {\cal D}\Phi]_{mag}
\, e^{ -S_e[A]-S_m[\Phi]}\; ,
\label{det}
\end{equation}
where the path integral is carried over fields $A_\mu$, $\Phi_\mu$
satisfying the condition (\ref{bc}).
Note that the original constraints yield boundary conditions corresponding 
to a perfect conductor of magnetic charges, while the field strength receives
contributions from both `electric monopole' ($A$) and `magnetic
monopole' ($\Phi$) sectors.

\section{Analysis of the gluon contribution \\ to the Casimir bag energy}

For the case of pure $SU(3)$~\footnote{The effect of dynamical fermions on field strength correlators 
can be studied on the lattice, see ref.~\cite{DiG}. The model parameters get renormalized, however, our general analysis is left unchanged.}, typical ansatze for $D$ and $D_1$ are
(see~ref. \cite{Otto}):
\begin{equation}
D(-z^2)=\frac{G_2}{24}\frac{27}{64}a^{-2}\int d^4k\, \frac{k^2}{\big(
k^2+(3\pi/8 a)^2 \big)^4}\, \exp(ikz) \,,
\label{iv}
\end{equation}
and
\begin{equation}
D_1(-z^2)=(z^2)^{-2}\int_0^{z^2} dv 2v D(-v) \;.
\end{equation}
We note that for large $z^2$, $D\sim exp \left(-\frac{3\pi|z|}{8a} \right)$ and,
\begin{equation}
D_1(-z^2)\sim (z^2)^{-2}.
\label{asy}
\end{equation}
Although a pure $SU(3)$ lattice calculation fixes the parameters to
$\kappa=0.74$, $a=0.35~fm$ and $G_2=(496~MeV)^4$, see \cite{latt,Otto} 
and references therein, it is interesting to study what happens when 
$\kappa$ goes from $0$ to $1$. 

When $\kappa=0$, there is just one propagating field,
\begin{equation}
\langle A_\mu (x) A_\nu (y) \rangle = -\tilde{H}\delta_{\mu \nu}\;,
\end{equation}
the leading contribution to $D_1$ is perturbative, $D_1\sim 1/z^4$,
and from (\ref{ident}) we see that $H\sim 1/z^2$. Thus, the leading
behaviour of the $A_\mu$ propagator is $\sim (1/k^2) \delta_{\mu
\nu}$. As this corresponds to the usual behaviour of $QED(4)$, the
Casimir effect at $\kappa=0$ is expected to be repulsive for a bag,
and after multiplying by $N=8$, $E_c \approx +0.7/R$. 

Then, for $\kappa$ close to zero, the usual result for the simplified
model where the Casimir bag energy is computed considering the gluon
modes as photon-like fields is recovered. Of course, the lattice fitting
value $\kappa=0.74$ is far from that regime, and it comes as no
surprise that the Casimir bag energy for the simplified model, after adding
contributions coming from center of mass fluctuations, is far from the
required value to adjust hadrons. 

Since $\kappa$ is in fact closer to one than to zero, let us analyze
the opposite regime, when $\kappa=1$. The field propagators are now given
by,
\begin{equation}
\langle A_\mu (x) A_\nu (y) \rangle =
\langle \Phi_\mu (x) \Phi_\nu (y) \rangle 
=-\tilde{D} \frac{\delta_{\mu \nu}}{k^2}\; .
\end{equation}
We can also write,
\begin{equation}
S_e[A]+S_m[\Phi]=\frac{1}{2}
\left( \begin{array}{cc}
A_\mu & \Phi_\mu \end{array} \right)
\delta_{\mu \nu}(\partial^2/{\tilde D})\left( \begin{array}{cc}
1 & 0  \\ 0 & 1
\end{array} \right)
\left( \begin{array}{c}
A_\nu \\
\Phi_\nu \end{array} \right)
\label{SeSm}
\end{equation}
(where in ${\tilde D}$ we have replaced $k^2$ by $-\partial^2$).

Now, we note that (\ref{iv}) implies,
\begin{equation}
\tilde{D}\propto \frac{k^2}{\big( k^2+\Lambda^2\big)^4}
\makebox[.5in]{,}
\Lambda=(3\pi/8a) \; ,
\label{DL}
\end{equation}
so that the kinetic operator in (\ref{SeSm}) is proportional to,
\begin{equation}
\delta_{\mu \nu}\, I_{2\times 2}\, (-\partial^2 +\Lambda^2)^4\; ,
\end{equation}
and the Casimir energy implied by (\ref{det}) is four times the one
for a kinetic operator,
\begin{equation}
\delta_{\mu \nu}\, I_{2\times 2}\, (-\partial^2 +\Lambda^2)\;.
\end{equation}

We see that the $A$ and $\Phi$ propagators are both {\em infrared
suppressed}, since the associated modes are in fact massive. The
lattice adjusted value of $a=0.35~fm$, corresponds to $\Lambda \sim
660~ Mev^{-1}$.  Casimir energies associated with massive modes at
this scale, have been precisely discussed in refs.~\cite{IRmod,bordag}, and are
strongly suppressed, for a typical hadron radius, when compared with
contributions of order $\sim 1/R$. Typical values of $R$ for the
smaller hadrons are $\sim 2/3 ~fm$, $\sim 1~fm$ ($1~fm\approx
\frac{1}{200} Mev^{-1}$).

It is interesting to note that the field representation we have obtained
here is quite similar to the simple infrared modified bag model
introduced in ref.~\cite{IRmod}. In that reference, information associated
with a mass scale $\Lambda$ was encoded in a quadratic Abelian field
theory, containing an effective gluon propagator, thus discussing
possible effects of nonperturbative infrared physics on
Casimir energy calculations.

With regard to the equivalent field theory discussed in the present
work, three important remarks are in order: i) the field theory has
been {\it derived} rather than assumed: its quadratic nature is a
direct consequence of the Gaussian dominance in shifted field strength
correlators, the `hard' part of the calculation, namely, the
generally nonlocal kernels in the kinetic terms, is borrowed from
lattice data; ii) as it is based on field strength correlators, the
formulation is SU(3) gauge invariant, as well as the defined Casimir energy
and the nonperturbative information encoded in the kernels; iii) It
contains a vector and a pseudovector field, representing electric and
magnetic monopole sectors; the field strength receives contributions
from both.

For the lattice value of $\kappa=0.74$, the Casimir effect will be
presumably closer to what happens at $\kappa=1$, where the effect is
suppressed with respect to contributions of order $\sim1/R$, than to
the value of $\sim +0.7/R$, at $\kappa=0$, implied by ``photon-like''
gluons. So that the $R^{-1}$ term in the bag energy is expected to be
attractive, dominated by center of mass fluctuations $-Z_{cm}/R$,
$Z_{cm}\sim 1$.

\section{Conclusions}\label{sec:conclusions}
In order to fit properly hadron spectroscopy, the MIT bag model
requires a term in the bag energy of the form $-Z/R$, with $Z$ of
order one.  As it is well known, when Casimir energy calculations are
done by means of the crudest simplification, considering `photon-like'
gluons, the effect turns out to be $\sim +0.7/R$, with the wrong sign.
Improvement on this disappointing result has been, for many years,
blocked by the fact that a full nonabelian calculation would be extremely
difficult to perform.

In this article, we have introduced a middle way to approach the
problem, namely, to take the most relevant nonabelian effects into
account.

Using a technique based on auxiliary fields living on the bag
boundary, to impose the bag constraints, we have been able to express
the Casimir bag energy in terms of shifted field strength correlators
in $SU(3)$. Of course, the obtained Casimir energy representation is
invariant under $SU(3)$ gauge transformations.

As lattice simulations show, the shifted field strength correlators
possess Gaussian dominance and can be parametrized in terms of two
structure functions fitted with the lattice data. This is the so
called stochastic vacuum model of QCD and it consistently describes
short distance perturbative aspects, such as asymptotic freedom, as
well as large distance nonperturbative ones, such as
confinement. Therefore, it is a natural approximation scheme to
discuss whether the simple model that treats gluon modes as free
Abelian massless fields can be applied, and if not, what are the
consequences of taking nonperturbative effects into account.

Gaussian dominance enables the introduction of an equivalent quadratic
field representation in terms of vectorial and pseudovectorial
effective fields, representing electric and magnetic monopole degrees
of freedom, respectively.  This representation contains two generally
nonlocal kernels, whose typical ansatze contains a $\kappa$ parameter
weighting the nonabelian nature of gluon fields. For $\kappa$ close to
zero, the effective model is dominated by perturbative physics, so
that in this case a ``photon-like'' gluon model would be obtained,
together with the associated Casimir energy, $\sim +0.7/R$.

However, the lattice adjusted value is $\kappa=0.74$, closer to
$\kappa=1$, a value where the equivalent Gaussian fields have a mass
gap $\Lambda=660~Mev$ and the Casimir energy for typical hadron radius
is strongly suppressed. Therefore, the Casimir energy of pure $SU(3)$ theory is
expected to be suppressed, when compared with attractive center of
mass contributions, $-Z_{cm}/R$, $Z_{cm}$ of order one. 
An important point is that the infrared suppression of propagators,
in our equivalent field representation, is SU(3) gauge invariant:
the fields represent effective modes with a nonperturbative gap, 
associated with SU(3) gauge invariant structure functions of shifted field 
strength correlators.

The predicted bag energy term is, therefore, $-Z/R$, 
$Z\sim Z_{cm}\sim 1$. To the best of our knowledge, this is the first controlled 
estimation of the $1/R$ term in the bag energy, and it turns out to comply with
hadron phenomenology.

\section*{Acknowledgements}
The Conselho Nacional de Desenvolvimento Cient\'{\i}fico e
Tecnol\'{o}gico (CNPq) and the Funda{\c {c}}{\~{a}}o de Amparo
{\`{a}} Pesquisa do Estado do Rio de Janeiro (FAPERJ) are acknowledged for
the financial support. C.\ D.\ F.\ acknowledges support by CONICET and
ANPCyT (Argentina). 

\end{document}